\documentclass[aps,prl,superscriptaddress,twocolumn,longbibliography,floatfix]{revtex4-2}

\usepackage[pdftex]{graphicx}
\usepackage{amsmath,amsfonts,amsbsy,amssymb,amsthm,mathtools}
\usepackage{mathrsfs}
\usepackage{ulem,color}
\usepackage{enumitem}
\usepackage{siunitx}
\usepackage{amssymb,amsmath,stmaryrd}
\usepackage{ulem,color}
\usepackage{MnSymbol}
\usepackage{environ}
\usepackage{xcolor}
\usepackage{stmaryrd}
\usepackage{multirow}
\usepackage{float}
\usepackage{gensymb}

\usepackage{tikz}
\usetikzlibrary{quantikz}
\usetikzlibrary{automata,arrows,positioning,calc}
\usetikzlibrary{arrows.meta,positioning}

\graphicspath{{./images/},{./imagesAppendix/}}

\newcommand{\prlsection}[1]{{\em {#1}.---}}

\usepackage{pgfplots}
\usepackage{pstool}
\usepackage{tikzsymbols}
\usetikzlibrary{plotmarks,spy}
\usepgfplotslibrary{external} %
\theoremstyle{definition}
\newtheorem{definition}{Definition}

\newcommand{\codes}{\mathcal{C}}
\newcommand{\rlxn}{\mathcal{P}}
\newcommand{\I}{\mathcal{I}}

\newcommand{\field}{\mathbb{F}}
\newcommand{\Rspace}{\mathbb{R}}

\newcommand{\K}{\mathcal{K}}
\newcommand{\M}{\mathcal{M}}
\newcommand{\y}{\mathbf{y}}
\DeclareMathOperator{\rank}{rank}

\usepackage[colorlinks = true, citecolor=blue, linkcolor=blue, urlcolor=blue]{hyperref}
\usepackage{mwe}
\usepackage{tikz}
\usetikzlibrary{quantikz}

\newcommand{\tr}{\textup{Tr}}
\renewcommand{\>}{\rangle}

\newcommand{\N}{{\mathbb{N}}} %
\newcommand{\R}{{\mathbb{R}}} %
\newcommand{\E}{{\mathbb{E}}}
\newcommand{\F}{{\mathbb{F}}}
\newcommand{\Z}{{\mathbb{Z}}}

\newcommand{\poly}{{\textup{Poly}}}
\newcommand{\sos}{{\textup{SOS}}}

\newcommand{\CQT}{Centre for Quantum Technologies, National University of Singapore, 3 Science Drive 2, Singapore 117543.\looseness=-1}
\newcommand{\NTU}{Nanyang Quantum Hub, School of Physical and Mathematical Sciences, Nanyang Technological University, Singapore 639673.\looseness=-1}
\newcommand{\ihpc}{Institute of High Performance Computing (IHPC), Agency for Science, Technology and Research (A*STAR), 1 Fusionopolis Way, $\#$16-16 Connexis, Singapore 138632, Republic of Singapore}
\newcommand{\qinc}{
Quantum Innovation Centre (Q.InC), Agency for Science Technology and Research (A*STAR), 2 Fusionopolis Way, Innovis $\#$08-03, Singapore 138634, Republic of Singapore }

\newcommand{\VT}{Department of Computer Science, Virginia Polytechnic Institute and State University, Blacksburg, Virginia 24061, USA.\looseness=-1}
\newcommand{\CSRU}{Cryptology and Security Research Unit, Indian Statistical Institute, Kolkata 700108, India.}
\newcommand{\ECSU}{Electronics and Communication Sciences Unit, Indian Statistical Institute, Kolkata 700108, India.}

\definecolor{THc}{rgb}{0.9,0.3,0.2}

\newcommand{\idg}[1]{{\bfseries #1)}}

\newcommand{\subfigimg}[3][,]{%
	\setbox1=\hbox{\includegraphics[#1]{#3}}%
	\leavevmode\rlap{\usebox1}%
	\rlap{\hspace*{2pt}\raisebox{\dimexpr\ht1-0.5\baselineskip}{{\bfseries \large\textsf{#2}}}}%
	\phantom{\usebox1}%
}

\newcommand{\SM}{SM}

\begin{document}

\normalem
\newlength\figHeight 
\newlength\figWidth 

\title{Hierarchical quantum decoders}

\author{Nirupam Basak}
\email{nirupambasak2020@iitkalumni.org}
\affiliation{\CSRU}

\author{Ankith Mohan}
\email{ankithmo@vt.edu}
\affiliation{\VT}

\author{Andrew Tanggara}
\email{andrew.tanggara@gmail.com}
\affiliation{\CQT}
\affiliation{\NTU}

\author{Tobias Haug}
\email{tobias.haug@u.nus.edu}
\affiliation{Quantum Research Center, Technology Innovation Institute, Abu Dhabi, UAE}

\author{Goutam Paul}
\email{goutam.paul@isical.ac.in}
\affiliation{\ECSU}
\affiliation{\CSRU}

\author{Kishor Bharti}
\email{kishor.bharti1@gmail.com}
\affiliation{\qinc}
\affiliation{\ihpc}

\date{\today}

\begin{abstract}
Decoders are a critical component of fault-tolerant quantum computing. They must identify errors based on syndrome measurements to correct quantum states.  While finding the optimal correction is NP-hard and thus extremely difficult, approximate decoders with faster runtime often rely on uncontrolled heuristics.  In this work, we propose a family of hierarchical quantum decoders with a tunable trade-off between speed and accuracy while retaining guarantees of optimality. 
We use the Lasserre Sum-of-Squares (SOS) hierarchy from optimization theory to relax the decoding problem. This approach creates a sequence of Semidefinite Programs (SDPs). Lower levels of the hierarchy are faster but approximate, while higher levels are slower but more accurate. We demonstrate that even low levels of this hierarchy significantly outperform standard Linear Programming relaxations. Our results on rotated surface codes and honeycomb color codes show that the SOS decoder approaches the performance of exact decoding. We find that Levels 2 and 3 of our hierarchy perform nearly as well as the exact solver. %
We analyze the convergence using rank-loop criteria and compare the method against other relaxation schemes. This work bridges the gap between fast heuristics and rigorous optimal decoding.
\end{abstract}

\maketitle

 \let\oldaddcontentsline\addcontentsline%
\renewcommand{\addcontentsline}[3]{}%

\prlsection{Introduction}
Although quantum computers have the potential to outperform classical computers, they are fragile to noise due to errors from interaction with the environment~\cite{zurek2003decoherence, preskill2018quantum}. To build reliable computers, we must correct these errors using Quantum Error Correction (QEC)~\cite{shor1995scheme, steane1996error, gottesman1997stabilizer, knill1997theory, fowler2012surface, aharonov1997fault, gottesman2002introduction, lidar2013quantum}. In QEC, we spread information across many physical qubits~\cite{bennett1996mixed, Kitaev1997quantum, Kitaev2003fault, fletcher2007optimum, fletcher2008structured, chao2018quantum, zurek1991decoherence, lloyd1993potentially, arute2019quantum, monroe2014large}. When errors occur, stabilizer measurements can generate specific patterns called syndromes. The process of identifying the actual error from the syndrome is called decoding. The performance of a quantum computer depending heavily on the decoder~\cite{dennis2002topological, fowler2012surface, terhal2015quantum}. Decoding involves finding the most probable error consistent with the syndrome, which is called Maximum Likelihood Decoding (MLD)~\cite{poulin2006optimal, ferris2014tensor}.
However, solving MLD exactly is very difficult, in fact it is an NP-hard problem~\cite{kuo2012hardness, hsieh2011np, berlekamp2003inherent}. 

Because exact decoding is hard, researchers often use heuristic algorithms that approximate MLD~\cite{bravyi2024highthreshold, poulin2008iterative, mackay2003information, duclos2010fast, varsamopoulos2017decoding, Gicev2023scalablefast, varbanov2025neural, torlai2017neural, hutter2014efficient, kubica2019cellular, demarti2024decoding, daskalakis2008probabilistic, li2018lp, javed2024low}. Examples include Minimum Weight Perfect Matching (MWPM)~\cite{wu2023fusion, fowler2015minimum, fowler2012towards, wang2003confinement} and Union-Find~\cite{das2022afs, skoric2023parallel, tan2023scalable, liyanage2023scalable, wu2022interpretation}. Although these are fast and scalable, their decoding accuracy can be limited, leading to lower error thresholds than MLD. On the other hand, exact solvers like Mixed Integer Programming (MIP) are optimal but too slow for large systems~\cite{Vanderbei2020integer, feldman2003decoding, feldman2005using}. Thus, there is a need for a framework that connects these two extremes.

We propose Hierarchical Quantum Decoders that allow a tunable trade-off between accuracy and speed. Our approach is based on the Lasserre hierarchy~\cite{lasserre2001global, lasserre2002explicit} or analogously the Sum-of-Squares (SOS) hierarchy~\cite{parrilo2003semidefinite, kim2005generalized} from optimization theory. This method tackle the hard discrete problem through a sequence of semidefinite programming relaxations.
Semidefinite programming (SDP) deals with optimizing a linear function of positive semidefinite matrices under a set of linear matrix constraints~\cite{tunccel2016polyhedral} (see End Matter for a formal definition).
Level 1 is a basic relaxation. It is computationally cheaper but less accurate~\cite{josz2018lasserre, burer2002rank, campos2023partial, sinjorgo2024solving, ghazi2017lp}. As we go to higher levels, the decoder becomes more accurate, accounting for complex correlations between errors. For a problem with $n$ variables, the $n$-th level of the hierarchy is sufficient to %
find the exact solution~\cite{lasserre2001global, lasserre2002explicit, navascues2008convergent}.

Our approach is rigorous and does not rely on ad-hoc rules. We test our decoder on the rotated surface code~\cite{bombin2007optimal} and the honeycomb color code~\cite{bombin2006topological}, comparing it to an exact MIP solver as reference. We observe that level 2 and 3 of our hierarchy achieve performance comparable to that of exact solvers, providing a tunable balance between computational speed and solution accuracy. 

\prlsection{Decoding problem}
When decoding quantum error-correcting codes, the standard approach is to find the most likely error given the measured syndromes~\cite{demarti2024decoding}. In particular, the code is subject to a configuration of $n$ different errors $e\in\F_2^n$ that matches the outcome of $v$ syndrome measurements $s\in\F_2^v$. 
The syndromes are described by the parity check matrix $H\in\F_2^{v\times n}$, with $s=He$.
We assume that the $i$-th error occurs with probability $p_i$, independent of other errors.
Then, the probability of error configuration $e$ is given by
\begin{equation}
    \Pr[e] = \prod_{i=1}^n (1-p_i)^{1-e_i} p_i^{e_i} = \Big(\prod_{i=1}^n (1-p_i)\Big) \prod_{i=1}^n \Big(\frac{p_i}{1-p_i}\Big)^{e_i}\,.
\end{equation}
Then, we find the most likely error, i.e., perform MLD
\begin{equation}\label{eqn:MLE_problem}
\begin{gathered}
    e^* = \arg\min_e \sum_{i=1}^n e_i\gamma_i \\
    \mathrm{s.t.}\quad
    He=s \quad\text{and}\quad
    e_i\in\F_2 ,\ i \in \{1, \dots, n\} \;,
\end{gathered}
\end{equation}
where we define weights $\gamma_i=\log \frac{1-p_i}{p_i}$. %
Note that MLD is NP-hard in general~\cite{hsieh2011np, berlekamp2003inherent}.

\prlsection{Sum-of-squares hierarchy}
We now show that MLD~\eqref{eqn:MLE_problem} can be equivalently formulated as a polynomial optimization problem.
This MLD polynomial optimization problem can then be relaxed to a hierarchy of SDPs that converges to the original problem. 
First we will define some notations, derive the particular form of polynomial optimization called the sum-of-squares (SOS) problem, and show that MLD can be solved using SOS hierarchy in Eq.~\eqref{eqn:SOS}.

Let us denote polynomial $f\in\R[x_1,\dots,x_n]$ over $n$ variables $x:=x_1,\dots,x_n$ as $f(x) = \sum_\alpha f_\alpha x^\alpha$ where we denote the monomial %
$x^{\alpha} \coloneqq x_1^{\alpha_1}\dots x_n^{\alpha_n}$ for $\alpha\in\Z_{\geq0}^n$, the set of length-$n$ strings of non-negative integers.
The total degree of $f$ is given by $\deg f = \max_\alpha \sum_{i=1}^n \alpha_i$.
The set of polynomials $f\in\R[x_1,\dots,x_n]$ with total degree of at most $t$ is denoted as $\poly_{n,t}$, whereas their corresponding set of all exponents is given by  $\Delta(n,t) = \{\alpha\in\Z_{\geq0}^n : \sum_{j=1}^n \alpha_j\leq t\}$.

A general polynomial optimization problem for objective polynomial $f\in\R[x_1,\dots,x_n]$ and $m$ constraint polynomials $\{g_i\}_{i=1}^m\subseteq\R[x_1,\dots,x_n]$ is given by
\begin{equation}\label{eqn:polynomial_optimization}
\begin{gathered}
    \min_{x\in\R^n} f(x) \\
    \mathrm{s.t.}\quad
    g_i(x)\geq0 ,\;\ i \in \{1, \cdots, m\},
\end{gathered}
\end{equation}
where we can define the feasible set of solutions as the semi-algebraic set $K = \{x\in\R^n : g_i(x)\geq0, \;\ i \in \{1, \cdots, m\}\}$.
Note that MLD~\eqref{eqn:MLE_problem} can be formulated as a polynomial optimization with objective polynomial $f(e) = \sum_{i=1}^n e_i\gamma_i$ and constraint polynomials $g_i(e) = e_i^2-e_i = 0$ (which is equivalent to $e_i\in\F_2$) for $i\in\{1,\dots,n\}$ and $H_je - s_j = 0$ for $j\in\{1,\dots,v\}$ where $H_j$ is the $j$-th row of $H$, giving us the following \textit{MLD polynomial optimization problem}
\begin{equation}\label{eqn:MLE_polynomial_optimization}
\begin{gathered}
    \min_e f(e) = \sum_{i=1}^n e_i\gamma_i \\
    \mathrm{s.t.}\quad
    h_j(e) = H_je - s_j = 0 \;,\; j\in\{1,\dots,v\} \;, \\
    g_i(e) = e_i^2-e_i = 0 \;,\; i\in\{1,\dots,n\} \;.
\end{gathered}
\end{equation}
Thus, in this case the number of constraints is $m=n+v$.

Polynomial $f\in\R[x_1,\dots,x_n]$ is a \textit{sum-of-squares} (SOS) polynomial if there exist polynomials $\{f_j\}_{j=1}^w$ such that $f=\sum_{j=1}^w f_j^2$.
The set of all $n$-variables sum-of-squares polynomials of total degree at most $t$ 
is denoted as $\poly_{n,t}^\sos$.
This allow us now to define the relaxations of the MLD polynomial optimization problem~\eqref{eqn:MLE_polynomial_optimization}.
First for hierarchy level $\ell\in\Z_{\geq1}$, we consider its Lagrangian 
\begin{equation}
\begin{gathered}
    L(x,q,q') = f(x) - \sum_{i=1}^n q_i(x) g_i(x) - \sum_{j=1}^v q_j'(x) h_j(x) \\
    \text{s.t.}\quad
    q_i \in \poly_{n,\ell+1}^\sos \quad\text{and}\quad
    q_j' \in \poly_{n,\ell}^\sos
\end{gathered}
\end{equation}
where $q=(q_1,\dots,q_n)$ and $q'=(q_1',\dots,q_v')$.

Then the SDP to solve the $\ell$-th level SOS hierarchy for the MLD polynomial optimization problem is given by %
\begin{equation}\label{eqn:SOS}
\begin{gathered}
    \max_{\lambda,q,q'} \lambda \\
    \text{s.t.}\quad
    L(x,q,q') - \lambda \geq0 \;,\\
    q_i \in \poly_{n,\ell+1}^\sos \quad\text{and}\quad
    q_j' \in \poly_{n,\ell}^\sos \;.
\end{gathered}
\end{equation}
It is known that the optimal value $\lambda_\ell^*$ of the $\ell$-th level SOS hierarchy converges to the optimal value $f(x^*)$ of its original polynomial optimization problem as $\ell\rightarrow\infty$ (see \cite[Section 4.2]{waki2006sums}, \cite[Theorem 4.1]{kim2005generalized}, \cite[Section 5]{Sagnol2019}).

Whether the $\ell$-th level of the SOS hierarchy has converged to the optimal value $f(x^*)$ of the original problem can be certified using the duality of SDP (see e.g.~\cite[Chapter 5]{boyd2004convex}).
It is known that for any $\ell$, the $\ell$-th level SOS in Eq.~\eqref{eqn:SOS} is \textit{dual} to the $\ell$-th level of the \textit{Lasserre} SDP hierarchy (see Definition~\ref{def:lasserre_hierarchy} in the \SM{}~\ref{sec:Lasserre}, also~\cite[Section 5]{Sagnol2019} and~\cite[Chapter 5]{lasserre2009moments}).
At $\ell$-th level Lasserre hierarchy, the positive semidefinite matrix $M_\ell$ indexed by monomials of degree $\leq\ell$ is called the $\ell$-th level \textit{moment matrix}~(see End Matter 
or \cite[Proposition 9]{Sagnol2019}). %
This is an $\binom{n+\ell}{\ell}\times\binom{n+\ell}{\ell}$ matrix. 
Whenever the rank of $M_\ell$ and $M_{\ell+1}$ are equal, it is guaranteed that the optimal value of the $\ell$-th level SOS~\eqref{eqn:SOS} is equal to the optimal value $f(x^*)$ of the original problem~\cite[Theorem~3.6]{lasserre2002explicit}. The $\ell$-th level of such hierarchy can be solved in $n^{\mathcal{O}(\ell)}$-time~\cite{barak2011rounding, barak2012hypercontractivity, chlamtac2012convex, lauria2017tight}.

\prlsection{Numerics}
Now, we study the performance of our SOS decoder as given in Eq.~\eqref{eqn:SOS} numerically. As example, we study the rotated surface code~\cite{bombin2007optimal} for bit-flip noise on the data qubits, while we assume that the syndrome measurements are noise-free  (see \SM{}~\ref{sec:surface}).

We also study the logical error rate $p_\text{L}$ against various physical error rates $p$, hierarchy levels of SOS relaxation $\ell$ and code distances $d$. The results are benchmarked against MLD implemented via MIP, which provides exact optimal decoding with high computational cost. To carry out the numerical optimization, we use the MATLAB-based \texttt{SparsePOP} package~\cite{Waki2008Algorithm}. This tool automatically reduces the problem size, when possible, by exploiting structural sparsity~\cite{fukuda2001exploiting, kim2005generalized, kojima2005sparsity, waki2006sums, gatermann2004symmetry, riener2013exploiting}. 

\begin{figure}[htpb]
\centering
\subfigimg[width=0.6\columnwidth]{}{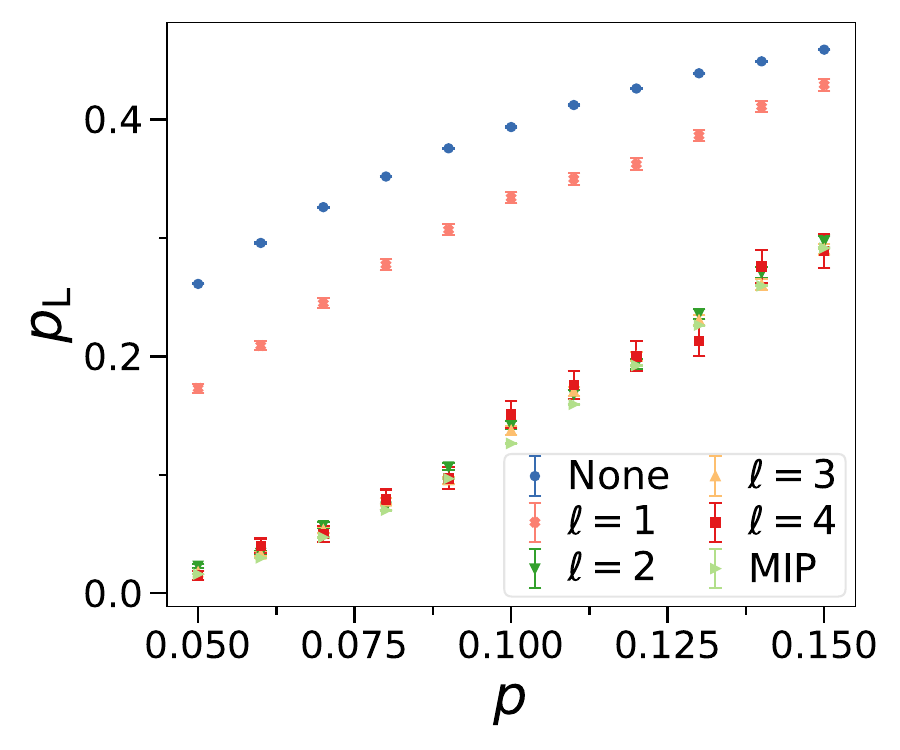}
\caption{Comparison of different SOS levels for decoding of surface code. We show logical error $p_\text{L}$ against physical error $p$ for rotated surface code for different types of decoders, where we fix distance $d=7$. We show no decoding at all, SOS decoder with level $\ell=1,\dots,4$ and the MIP decoder. %
}
\label{fig:surfacecode_comp}
\end{figure}

First, we study in Fig.~\ref{fig:surfacecode_comp} $p_\text{L}$ against $p$ for fixed code distance $d=7$.
Notably, we see that already level $\ell=1$ gives improved performance over no decoding at all. For $\ell\geq2$, we observe performance nearly as good as the optimal MIP decoding. 

\begin{figure}[htpb]
\centering
\subfigimg[width=0.5\columnwidth]{a}{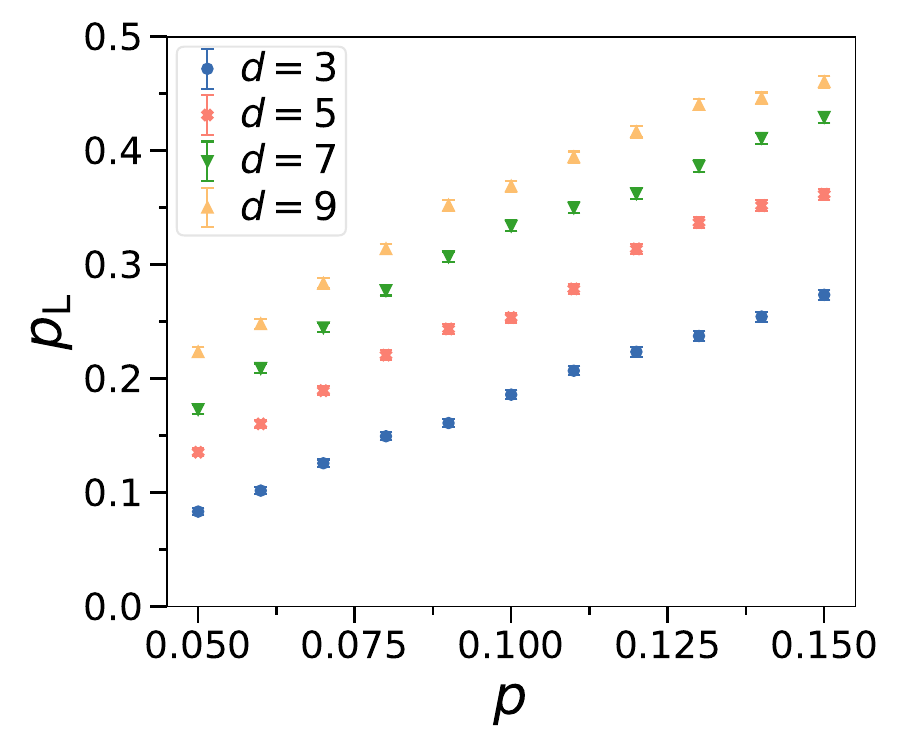}\hfill
\subfigimg[width=0.5\columnwidth]{b}{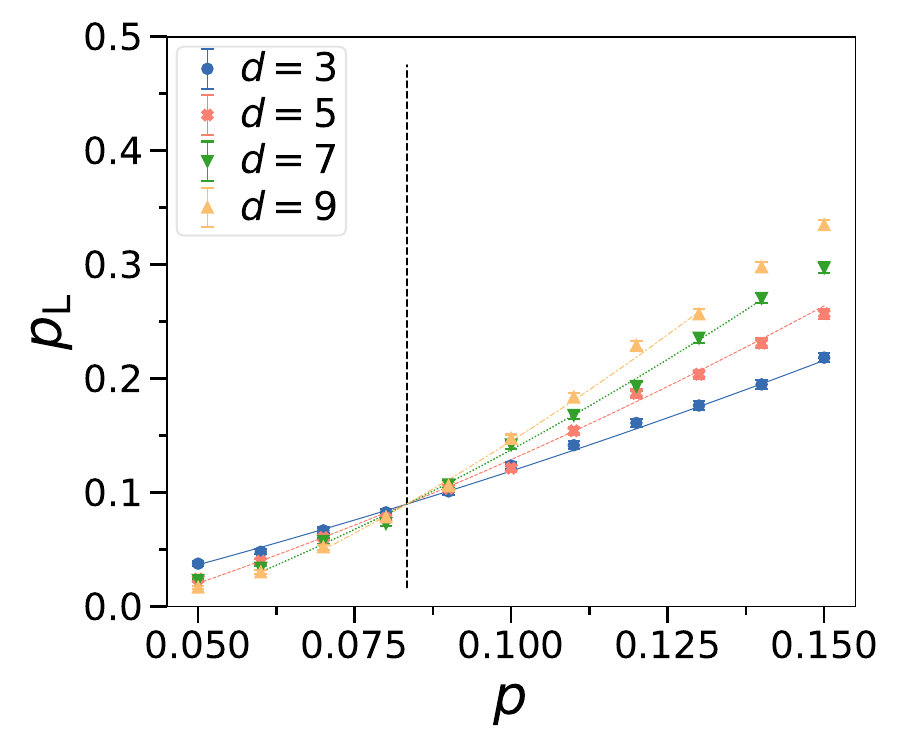}\\
\subfigimg[width=0.5\columnwidth]{c}{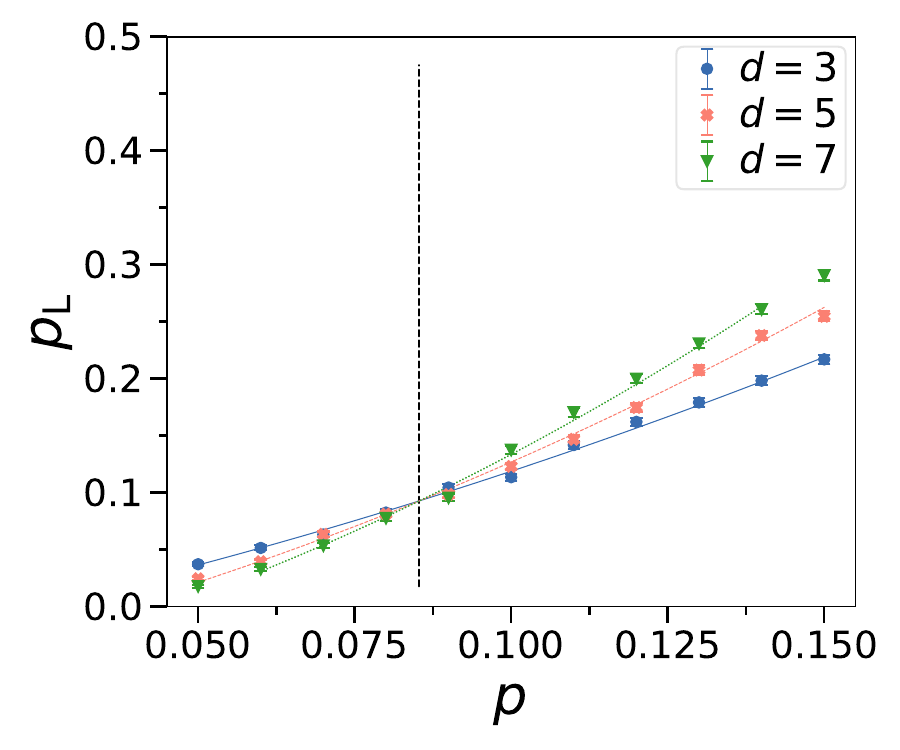}\hfill
\subfigimg[width=0.5\columnwidth]{d}{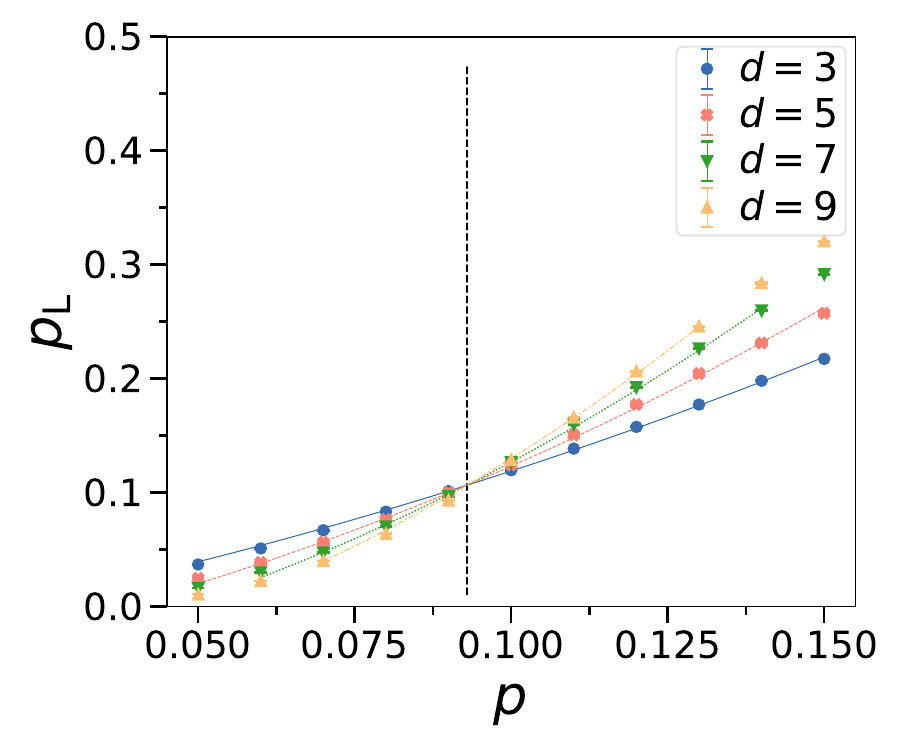}
\caption{Decoding rotated surface code. We show logical error $p_\text{L}$ against physical error $p$ for rotated surface code with distance $d\in\{2,5,7,9\}$. We show SOS hierarchy \idg{a} level 1, \idg{b} level 2, \idg{c} level 3 and \idg{d} MIP decoder. Dots are simulated data, while curves are fit with~\eqref{eq:fit_thr}.
Vertical dashed line shows the numerically fit of threshold $p_\text{th}=\{\text{N/A}, 0.083, 0.085, 0.093\}$.
}
\label{fig:surfacecode}
\end{figure}

Next, in Fig.~\ref{fig:surfacecode} we show $p_\text{L}$ for different $d$ for the SOS decoder at level $\ell\in\{1,2,3\}$ (Fig.~\ref{fig:surfacecode}a-c) and MIP decoder (Fig.~\ref{fig:surfacecode}d). %
We estimate the threshold $p_\text{th}$ by fitting the crossing point of the logical error rate curves of different $d$. Notably, we observe that the SOS decoder has a threshold for all $\ell\geq2$.
To determine the threshold, we fit with~\cite{chubb2021general}
\begin{equation}\label{eq:fit_thr}
    p_\text{L}(p,d)=f(d^{1/\nu}(p-p_\text{th})),
\end{equation}
where $f(x)=a+bx+cx^2$ is a second-order polynomial.
We observe that the threshold for SOS decoder is close to that of the MIP decoder, with improving threshold as $\ell$ increases.

\begin{table}[htpb]
\centering
\begin{tabular}{|c|c|c|c|}
\hline
\multicolumn{2}{|c|}{\textbf{Decoder}}& \textbf{Error} & \textbf{Rank} \\
\hline
\multirow{4}{*}{\textbf{SOS}}&\textbf{Level 2} & $0.0231(15)$ & $171(1)$ \\
\cline{2-4}
&\textbf{Level 3} & $0.0250(16)$ & $239(2)$ \\
\cline{2-4}
&\textbf{Level 4} & $0.0255(16)$ & $261(1)$ \\
\cline{2-4}
&\textbf{Level 5} & $0.0253(16)$ & $262(2)$ \\
\hline
\multicolumn{2}{|c|}{\textbf{MIP}} & $0.0248(2)$ &  \\
\hline
\end{tabular}
\caption{Logical error rates $p_L$ for different decoders, together with the rank of the moment matrix for the SOS decoder, are reported for the distance-5 surface code under bit-flip noise at a physical error rate $p=0.05$. The standard deviation is given in parentheses. The results indicate that the rank of the moment matrix stabilizes at level~5.
}
\label{tab:rank}
\end{table}

Finally, we compare in Table~\ref{tab:rank} logical error rates from the SOS and MIP decoders for the distance-5 surface code. As the hierarchy level increases, the SOS decoder produces steadily more accurate estimates, with logical error rates that approach those of the exact MIP decoder. We also show the rank of the moment matrix of the SOS decoder for different levels. The rank of the moment matrix increases with hierarchy level, and converges to its maximum rank at a particular critical level $\ell_\text{c}=5$. For $\ell\geq \ell_\text{c}$, the SOS decoder yields the optimal solution of the MLD problem. The logical error rates for distance-7 surface code obtained from SOS and MIP decoders are compared in Table~\ref{tab:surface} (see \SM{}~\ref{sec:surface}).
We also study the honeycomb color code~\cite{bombin2006topological} in \SM{}~\ref{sec:color}, where we observe similar behaviour in performance as with the surface code.

\prlsection{Discussion} %
We have introduced Hierarchical Quantum Decoders which offers a rigorous optimization theory approach to quantum error correction. We showed that the decoding problem can be solved by a hierarchy of SDPs. The results are promising. While the basic relaxation (level 1) is insufficient, level 2 and 3 perform nearly as well the MIP decoder, approaching the logical error reduction of the optimal MLD. Notably, we observe good performance also for the color code described in \SM{}~\ref{sec:color}, which is known to be challenging to decode~\cite{duclos2010fast}.

In our numerics, we observe that the simplest relaxation (level 1) suffers from poor performance, which we explain with the fact that error variables can assume any real value between 0 and 1. 
A similar problem has been noted in LP relaxations of classical decoding. Here, ``pseudocodewords'' appear, which are are fractional solutions that look like valid errors but are physically impossible (e.g., 0.5 error)~\cite{feldman2003decoding}. As classical decoding is equivalent to MLD (see \SM{}~\ref{sec:MLD}), we believe these pseudocodewords also affect our level 1 decoder.   %
Indeed, Gu and Soleimanifar~\cite{gu2025power} recently characterized these fractional solutions in quantum LDPC codes, identifying that they arise from specific constant-weight error patterns related to cycles in the Tanner graph. While they address this by employing Ordered Statistics Decoding (OSD) as a post-processing step, we find that moving to Level 2 of the hierarchy effectively suppresses these pseudocodewords by enforcing higher-order constraints, forcing variables to be either 0 or 1. This removes the pseudocodewords and leads to a finite threshold.

We also highlight that beyond its tunable level structure, the Lasserre hierarchy provides another unique feature: a certificate of optimality. This is known as the rank-loop condition (or flat extension)~\cite{laurent2009generalized}. When we solve the SDP at a certain level, we get a moment matrix. If the rank of this matrix satisfies a specific condition (rank stabilizes), we know that the solution must be exact~\cite[Theorem~3.6]{lasserre2002explicit}.
 In our simulations, we often observe this rank condition at level~5, which is shown in Table~\ref{tab:rank}.

We note that there are other hierarchy methods beyond the one we studied, such as Sherali-Adams (SA) and Lovasz-Schrijver (LS) (\SM{}~\ref{sec:lift-and-project}). 
For 0/1 programs starting from a common linear relaxation $\rlxn$, at any level $\ell$ we have:
\begin{equation}
LS^{(\ell)}(\rlxn) \subseteq SA^{(\ell)}(\rlxn) \subseteq Las^{(\ell)}(\rlxn),
\end{equation}
in the lifted space.
After projection onto the original space, the Lasserre hierarchy yields the tightest relaxation among these methods, though at a higher computational cost per level (see \SM{}~\ref{sec:comp_rlxn}).

Our work clarifies the landscape of decoders by providing a rigorous path from fast heuristics to exact solvers. While we note that solving SDPs is currently slower than other, less rigorous approximations such as graph matching,  our decoder has a unique feature among the decoding landscape: It offers a rigorous approach to tailoring the trade-off between accuracy and speed, with provable guarantees to optimality. Thus, 
for example, our solver is well suited for benchmarking small-size to medium-size codes and analyzing the properties of new codes where less rigorous heuristics might fail.

We also note that our method shares similarities with decoders that formulate MLD as quadratic unconstrained binary optimization (QUBO) problems, such as for coherent Ising machine~\cite{fujisaki2022practical,takada2024ising} or similarly for Max-SAT~\cite{berent2024decoding,noormandipour2024maxsat}  (see \SM{}~\ref{sec:MLE_QUBO}). However, while QUBO optimizers are usually heuristic, our approach offers systematic optimization with guaranteed convergence via SDPs.
Finally, our methods can be straightforwardly adapted to determine the code distance, which is also an NP-hard problem, via a hierarchy of SDPs.
Future work could improve the speed by exploiting the sparsity of the problem, potentially yielding high-accuracy decoders practical for larger systems.

\prlsection{Acknowledgements} 
AT is supported by the CQT PhD scholarship, the Google PhD fellowship program, and the CQT Young Researcher Career Development Grant. 
KB thanks HQCC WP 2.0 for financial support.
The authors thank Jamie Sikora and Shouzhen Gu for helpful discussions.

\bibliography{references}

\newpage

\let\addcontentsline\oldaddcontentsline

\onecolumngrid
\vspace{2cm}
\begin{center}

\textbf{\large End Matter}
\end{center}

\prlsection{Semidefinite programming} Semidefinite programming (SDP) deals with optimizing linear function of positive semidefinite matrices over affine constraints.
In the standard form, a SDP can be expressed as
\begin{equation*}
    \mathrm{minimize:}\; \left\{ \tr(C X)\ :\ \tr(A_i X) = b_i,\ i \in \{1, \cdots, m\},\ X \succcurlyeq 0 \right\},
\end{equation*}
where $C$ and $A_i$ are Hermitian matrices and $b_i$ are real.
The dual of SDP is
\begin{equation*}
    \mathrm{maximize:}\; \left\{ b^\top y\ :\ \sum_{i=1}^m y_i A_i \preccurlyeq C,\ y \in \mathbb{R}^m \ \right\}.
\end{equation*}

A simple example of a commonly occurring problem in quantum theory that can be expressed as an SDP is the ground state energy problem, i.e., finding the smallest eigenvalue of a Hamiltonian $H$ in a finite-dimensional Hilbert space.
This problem can be formulated as
\begin{equation*}
    \mathrm{minimize:}\; \left\{ \tr(H \rho)\ :\ \tr(\rho) = 1,\ \rho \succcurlyeq 0 \right\}. 
\end{equation*}

It is useful to cast problems as SDP as this allows to solve the problem either analytically or numerically.
Even when one cannot solve the problem exactly, SDPs can help in finding good bounds or approximations.

In quantum information, positive semidefinite matrices are foundational for describing quantum states, measurements, and channels.
Since these matrices are closed under addition and positive scalar multiplication, they form convex cones.
Density matrices, completely positive maps, separable operators, and entanglement witnesses are some common examples of cones of positive semidefinite matrices.
Entanglement detection, quantum state/channel discrimination, quantum non-local games, quantum channel capacity, and quantum marginal problem are some problems that can be  solved or approximated using SDP. Apart from quantum information, SDP is widely applied in varied fields such as control theory, combinatorial optimization, machine learning, and polynomial optimization. The reader is referred to~\cite{tunccel2016polyhedral} for a more detailed exposition.

\prlsection{The Lasserre Moment Hierarchy} Note that we formulate MLD as a Polynomial optimization problem (POP). To solve this POP (equation~\ref{eqn:MLE_polynomial_optimization}), we employ the Lasserre hierarchy (or Moment-SOS hierarchy), which relaxes the non-convex POP into a sequence of SDPs (see SM~\ref{sec:Lasserre} for details). The core idea is to lift the optimization variable from the vector $e$ to a probability measure $\mu$ supported on the feasible set $\K = \{e \in \R^n \mid h_j(e)=0, e_i^2-e_i=0\}$. Instead of finding the optimal point directly, we seek the moments of this measure. We define the moment sequence $\y = (y_\alpha)_{\alpha \in \N^n}$ where $y_\alpha = \int e^\alpha d\mu(e) = \mathbb{E}_\mu [e_1^{\alpha_1} \dots e_n^{\alpha_n}]$. In simple terms, suppose we do not know the exact error, but we have a distribution $\mu(e)$. The moments of this distribution are the expectation values of products of error variables: $y_\alpha = \E[e^\alpha] = \E[e_1^{\alpha_1} \dots e_n^{\alpha_n}]$. 

For the measure $\mu$ to exist, the sequence $\y$ must satisfy specific positivity conditions. We define the \textit{Moment Matrix} $M_\ell(\y)$ of level $\ell$, indexed by monomials up to degree $\ell$. The entries are given by:
\begin{equation}
    [M_\ell(\y)]_{\alpha, \beta} = y_{\alpha+\beta}, \quad |\alpha|, |\beta| \le \ell.
\end{equation}

For example, consider a simple system with 2 qubits. At Level 1 ($\ell=1$), we track moments up to degree 2. The basis is $\{1, e_1, e_2\}$, and the Moment Matrix is:
\begin{equation}
    \M_1 = \begin{pmatrix}
    1 & \langle e_1 \rangle & \langle e_2 \rangle \\
    \langle e_1 \rangle & \langle e_1^2 \rangle & \langle e_1 e_2 \rangle \\
    \langle e_2 \rangle & \langle e_1 e_2 \rangle & \langle e_2^2 \rangle
    \end{pmatrix}.
\end{equation}
The off-diagonal term $\langle e_1 e_2 \rangle$ represents the correlation between an error on qubit 1 and qubit 2. A necessary condition for $\y$ to represent a valid measure is $M_\ell(\y) \succeq 0$. 
At level $\ell$ of the hierarchy, we optimize over moment sequences up to degree $2\ell$:
\begin{align}
    \min_{\y} \quad & \sum_{i=1}^n \gamma_i y_{e_i} \\
    \text{s.t.} \quad & y_0 = 1, \nonumber \\
    & M_\ell(\y) \succeq 0, \nonumber \\
    & M_{\ell - d_j}(h_j \y) = 0 \quad \forall j. \nonumber
\end{align}
The linear constraints $M_{\ell - d_j}(h_j \y) = 0$ (where $d_j = \lceil \deg(h_j)/2 \rceil$) enforce the stabilizer conditions. In simple terms, at level $\ell$, the decoder is an SDP that optimizes the matrix $\M_\ell$ subject to three types of constraints:
\begin{enumerate}
    \item \textit{PSD Constraint ($\M_\ell \succeq 0$):} This ensures the moments come from a valid distribution. It forces correlations to satisfy inequalities like $\langle e_i^2 \rangle \langle e_j^2 \rangle \geq \langle e_i e_j \rangle^2$.
    \item \textit{Binary Consistency:} Since $e_i^2 = e_i$ for binary variables, we enforce linear constraints on the matrix entries, e.g., $\M_{\alpha, \beta} = \M_{\alpha', \beta'}$ if the underlying monomials simplify to the same expression. In the example above, we force $\langle e_1^2 \rangle = \langle e_1 \rangle$.
    \item \textit{Syndrome Consistency:} The physical parity checks $He=s$ are imposed as linear constraints on the rows of the matrix.
\end{enumerate}

The hierarchy can equivalently be viewed from the dual perspective. The dual to the moment minimization problem is the Sum-of-Squares (SOS) problem. We seek the largest lower bound $\lambda$ (see equation~\ref{eqn:SOS}) such that the shifted objective function can be decomposed as:
\begin{equation}
    E(e) - \lambda = \sigma_0(e) + \sum_j q_j(e) h_j(e) + \sum_i p_i(e) (e_i^2 - e_i),
\end{equation}
where $\sigma_0(e)$ is a sum-of-squares polynomial (guaranteeing non-negativity) and $q_j, p_i$ are arbitrary polynomial multipliers. If this decomposition exists, then clearly $E(e) \ge \lambda$ for all feasible $e$. The main text refers to this as the ``SOS hierarchy,'' while the numerical implementation often solves the primal moment problem. Strong duality holds for this hierarchy in our case, ensuring the optimal values coincide.

Unlike heuristic decoders, the Lasserre hierarchy provides a sufficient condition for exactness known as the \textit{rank-loop} or \textit{flat extension} property. Let $M_\ell(\y^*)$ be the optimal moment matrix at level $\ell$. If the rank stabilizes between levels, i.e.,
\begin{equation}
    \rank M_\ell(\y^*) = \rank M_{\ell-1}(\y^*),
\end{equation}
then the relaxation is exact.

The dimension of the moment matrix $M_\ell(\y)$ scales as $\binom{n+\ell}{\ell}$. For dense problems, this complexity limits $\ell$ to very small values. However, Quantum error correction codes exhibit \textit{correlative sparsity}: parity checks involve only local subsets of qubits. We utilize the sparse variant of the Lasserre hierarchy (as implemented in \texttt{SparsePOP}). The global PSD constraint $M_\ell(\y) \succeq 0$ is replaced by a set of smaller PSD constraints on sub-matrices corresponding to the maximal cliques of the correlative graph. This graph connects variables $e_i, e_j$ if they appear together in any term of the objective or constraints.

\newpage

\clearpage

\appendix
\onecolumngrid

\begin{center}

\textbf{\large Supplemental Material}
\end{center}

\setcounter{secnumdepth}{2}
\renewcommand{\thesection}{\Alph{section}}
\renewcommand{\thesubsection}{\arabic{subsection}}
\renewcommand*{\theHsection}{\thesection}
\renewcommand\appendixname{\SM{}}

In the Supplemental Material, we provide additional proofs and results to support the main text. %

\makeatletter
\@starttoc{toc}
\makeatother

\section{Lasserre hierarchy}\label{sec:Lasserre}
Here we discuss the Lasserre hierarchy, which is a dual of the SOS hierarchy (Eq.~\eqref{eqn:SOS}).
It is known (see e.g.~\cite[Theorem 6]{Sagnol2019}) that a polynomial $f$ is SOS if and only if there exists some positive semidefinite matrix $M=[M_{\alpha,\beta}]_{\alpha,\beta\in\Delta(n,t)}$ such that
\begin{equation}\label{eqn:SOS_PSD_correspondence}
    \sum_{\alpha,\beta\in\Delta(n,t):\alpha+\beta=\tau} M_{\alpha,\beta} = f_\tau
\end{equation}
for all $\tau\in\Delta(n,2t)$ (recall $\Delta(n,t) = \{\alpha\in\Z_{\geq0}^n : \sum_{j=1}^n \alpha_j\leq t\}$ as defined in the main text). %
By using a
column vector of monomials $v_x = (x^\alpha)_{\alpha\in\Delta(n,t)}$, we can obtain the SOS polynomial $f$ corresponding to positive semidefinite matrix $M=[M_{\alpha,\beta}]_{\alpha,\beta\in\Delta(n,t)}$ as
\begin{equation}
\begin{gathered}
    f(x) = v_x^\top M v_x = \sum_{\tau\in\Delta(n,2t)} f_\tau x^\tau \\
    \text{for}\\
    f_\tau = \sum_{\alpha,\beta : \alpha+\beta=\tau} M_{\alpha,\beta} \;.
\end{gathered}
\end{equation}

The optimizer obtained from the $\ell$-th level Lasserre hierarchy can be used to construct a matrix called the \textit{moment matrix}.
As mentioned in the the paragraph below Eq.~\eqref{eqn:SOS} (see main text), the rank of the moment matrix can be used as a certificate on whether the SOS hierarchy has converged to the original optimization problem.
The Lasserre hierarchy is obtained by formulating a polynomial optimization problem~\eqref{eqn:polynomial_optimization} as a \textit{moment problem}, which can be formulated as an SDP optimization problem.
Relaxations of this moment problem gives us the Lasserre hierarchy.

First, note that a polynomial optimization problem~\eqref{eqn:polynomial_optimization} can be written as a \textit{moment problem} over the semi-algebraic set $K = \{x\in\R^n : g_i(x)\geq0, \;\forall i\in[m]\} \subseteq\R^n$ for $[m]=\{1,\dots,m\}$ (defined by constraints $\{g_i\}_{i=1}^m$) as:
\begin{equation}\label{eqn:moment_problem1}
\begin{gathered}
    \min_{\mu\in\mathcal{M}^+(K)} \int_K f(x) \,d\mu(x)
\end{gathered}
\end{equation}
where the minimization is over all measures with support in $K$, i.e. $\mu(K)=1$, and $\mathcal{M}^+(K)$ the set of measures over the semi-algebraic set $K$:
\begin{equation}\label{eqn:measure_feasible_set}
    \mathcal{M}^+(K) = \Big\{ \mu\in\mathcal{M}^+(\R^n) : \mu(\R^n\backslash K)=0 \Big\} \;,
\end{equation}
where $\mathcal{M}^+(\R^n)$ is the set of all non-negative measures over $\R^n$.

For a measure $\mu\in\mathcal{M}^+(K)$ we have a \textit{moment sequence} $y=(y_\alpha)_{\alpha\in\Z_{\geq0}^n}$ with $y_{0^n} = \int_K x^{0^n} \,d\mu(x) = 1 = \mu(K) = \mu(\R^n)$.
Thus for degree-$t$ polynomial $f(x) = \sum_\alpha f_\alpha x^\alpha \in\R[x_1,\dots,x_n]$, we define
\begin{equation}
\begin{aligned}
    \<f,y\> & := \sum_{\alpha\in\Delta(n,t)} f_\alpha y_\alpha = 
    \int_K f(x) \,d\mu(x) \;,
\end{aligned}
\end{equation}
which allow us to rewrite Eq.~\eqref{eqn:moment_problem1} as
\begin{equation}\label{eqn:moment_problem2}
\begin{gathered}
    \min_{y\in \mathcal{M}_t^+(K)} \<f,y\> \\
    \textup{s.t.}\quad y_{0^n}=1 \;,
\end{gathered}
\end{equation}
where $\mathcal{M}_t^+(K)$ is the set containing all moment sequences $y = (\int_{\R^n} x^\alpha \,d\mu)_{\alpha\in\Delta(n,t)}$ with degree up to $t$ and $\mu\in\mathcal{M}^+(K)$. 

Now we would like to relax the problem in Eq.~\eqref{eqn:moment_problem2}.
First we need to define a \textit{localizing matrix} $M_r(gy)$ and a \textit{moment matrix} $M_r(y)$ for $r\in\N$, polynomial $g\in\poly_{n,2u}$, and moment sequence $y = (\int_K x^\alpha \,d\mu)_{\alpha\in\Z_{\geq0}^n}$ for $\mu\in\mathcal{M}^+(K)$. 
Their entries are indexed by polynomial degrees $\alpha,\beta\in\Delta(n,r)$, given by
\begin{equation}
    [M_r(gy)]_{\alpha,\beta} = \sum_{\gamma : |\gamma|\leq 2u} g_\gamma y_{\alpha+\beta+\gamma}
    \quad\text{and}\quad
    [M_r(y)]_{\alpha,\beta} = y_{\alpha+\beta} \;.
\end{equation}
Localizing matrix $M_r(gy)$ and moment matrix 
$M_r(y)$ are known to satisfy $M_r(y) \geq0$ and $M_r(g_i y) \geq0$ for all $i\in[m]$ and all $r\in\N$~\cite[Proposition 11]{Sagnol2019}.
This allows us to define the set of moment sequences
\begin{equation}
\begin{aligned}
    \mathcal{M}_{2r}^\mathrm{SDP}(K) &= \Big\{ y\in\R^{\chi(n,2r)} : M_r(y)\geq0 \;,\;\textup{and}\; M_{\xi_i}(g_iy)\geq0, \forall i\in[m] \Big\}
\end{aligned}
\end{equation}
which satisfy $\mathcal{M}_{2r}^+(K) \subseteq \mathcal{M}_{2r}^\mathrm{SDP}(K)$ (see~\cite[Proposition 11]{Sagnol2019}), where $\chi(n,t) = |\Delta(n,t)| = \binom{n+t}{t}$ is the size of the set of all possible degrees in $\Delta(n,t)$ and $\xi_i = r-\lceil\deg(g_i)/2\rceil$.
A stricter inclusion than $\mathcal{M}_{2r}^+(K) \subseteq \mathcal{M}_{2r}^\mathrm{SDP}(K)$ can be obtained by defining a sequence of sets that lies between $\mathcal{M}_{2r}^+(K)$ and $\mathcal{M}_{2r}^\mathrm{SDP}(K)$:
\begin{equation}
\begin{aligned}
    \mathcal{M}_{2r,\ell}^\mathrm{SDP}(K) &= \Big\{ y'\in\R^{\chi(n,2r)} : \exists y\in\R^{\chi(n,(2(r+\ell))} \;\textup{s.t.} \\
    &\quad y_\alpha'=y_\alpha, \forall|\alpha|\leq 2r\\
    &\quad\textup{and}\; M_{r+\ell}(y')\geq0 \\ &\quad\textup{and}\; M_{\xi_i+\ell}(g_iy)\geq0, \forall i\in[m] \Big\}
\end{aligned}
\end{equation}
for $\xi_i = r-\lceil\deg(g_i)/2\rceil$ and $\ell\in\N$. 
It can be shown that these sets satisfy the following sequence of inclusions~\cite[Corollary 12, Theorem 13]{Sagnol2019}
\begin{equation}
\begin{gathered}
    \mathcal{M}_{2r}^+(K) \subseteq\dots\subseteq \mathcal{M}_{2r,1}^\mathrm{SDP}(K) \subseteq \mathcal{M}_{2r,0}^\mathrm{SDP}(K) = \mathcal{M}_{2r}^\mathrm{SDP}(K) \\
    \text{and}\\
    \mathcal{M}_{2r}^+(K) = \bigcap_{\ell\in\N} \mathcal{M}_{2r,\ell}^\mathrm{SDP} \,,
\end{gathered}
\end{equation}
assuming that the Archimedean condition holds: There exists $R\in\R$ and SOS polynomials $\sigma_1,\dots\sigma_m\in\poly^\mathrm{SOS}$ such that $R^2-||x||^2 = \sum_{i=1}^m \sigma_i(x)g_i(x)$ for all $x\in K$.
Therefore, by optimizing over $\mathcal{M}_{2r,\ell}^\mathrm{SDP}(K)$ with increasing $\ell$ we are guaranteed to eventually converge to the optimal value $f(x^*)$ of the original polynomial optimization problem~\eqref{eqn:polynomial_optimization}.
This inclusions give us the Lasserre hierarchy.

\begin{definition}[{Lasserre hierarchy}]\label{def:lasserre_hierarchy}
    For polynomial optimization problem in Eq.~\eqref{eqn:polynomial_optimization} with objective polynomial $f\in\R[x_1,\dots,x_n]$ and constraint polynomials $\{g_i\}_{i=1}^m$, the $\ell$-th \textit{Lasserre hierarchy} problem is given by 
    \begin{equation}
    \begin{gathered}
        \min_{y\in\R^{\chi(n,2(r+\ell))}} \<f,y\> \\
        \textup{s.t.}\\
        y_0=1 \;,\\
        M_{r+\ell}(y)\geq0 \;,\\
        M_{\xi_i+\ell}(g_iy)\geq0, \forall i\in[m]
    \end{gathered}
    \end{equation}
    for $\xi_i = r-\lceil\deg(g_i)/2\rceil$ and $r=\max\{\lceil (\deg f)/2\rceil , u\}$ with $u=\max\limits_i \deg(g_i)/2$.
\end{definition}

\section{Rotated Surface Code}\label{sec:surface}

The surface code is a topological stabilizer code that protects quantum information using a two-dimensional lattice of qubits~\cite{Kitaev1997quantum, Kitaev2003fault}. Quantum information is preserved through measurements of two types of stabilizers: vertex or star operators, defined as products of Pauli-$X$ on qubits adjacent to a vertex in the lattice to detect $Z$-type errors, and plaquette or face operators, defined as products of Pauli-$Z$ around each face in the lattice to detect $X$-type errors~\cite{dennis2002topological, Kitaev2003fault, fowler2009high}.

The rotated surface code is a variant of the standard planar surface code in which qubits are placed on the vertices of a 2-dimensional square lattice rotated by $45^\degree$ and the stabilizer operators occupy alternating plaquettes in a checkerboard pattern~\cite{bombin2007optimal}. This reduces the number of physical qubits relative to the unrotated planar layout while preserving the code distance~\cite{orourke2025compare}. The distance-$d$ rotated surface code encodes one logical qubit with parameters $\llbracket d^2, 1, d\rrbracket$ on a $d\times d$ lattice.

The stabilizer group of the rotated surface code can be described formally as a CSS code generated by two sets of commuting operators. Let $F_X$ and $F_Z$ denote the sets of plaquettes hosting $X$-type and $Z$-type stabilizers, respectively, arranged in the checkerboard pattern. Then the stabilizers are written as
\[
S^X_f = \prod_{v\in\partial f} X_v \quad \text{for } f\in F_X, 
\;
S^Z_f = \prod_{v\in\partial f} Z_v \quad \text{for } f\in F_Z,
\]
where $\partial f$ denotes the set of vertices on the boundary of face $f$. The corresponding CSS parity-check matrices are
\[
H_X \in \{0,1\}^{|F_X|\times d^2}, 
\;
H_Z \in \{0,1\}^{|F_Z|\times d^2},
\]
where the entry $(H_X)_{f,v} = 1$ if $v\in \partial f$, and similarly for $H_Z$. These matrices satisfy $H_X H_Z^T = 0$, formalizing the commutation relations required for a CSS code. In the interior of the lattice, $|\partial f| = 4$, giving weight-4 checks, while boundaries yield weight-2 checks.

\begin{table}[htpb]
\centering
\begin{tabular}{|c|c|c|c|c|}
\hline
\multicolumn{2}{|c|}{\multirow{2}{*}{\textbf{Decoder}}}&\multicolumn{2}{c|}{\textbf{Error}}&\textbf{Rank of}\\
\cline{3-4}
\multicolumn{2}{|c|}{}& \textbf{$p=0.05$} & \textbf{$p=0.15$}&\textbf{moment matrix}\\
\hline

\multirow{4}{*}{\textbf{SOS}}&\textbf{Level 1} & $0.1731 (38)$ & $0.4294(49)$ & $228(27)$\\
\cline{2-5}
&\textbf{Level 2} 
& $0.0231(15)$ & $0.2973(38)$ & $1037(1)$\\
\cline{2-5}
&\textbf{Level 3} 
& $0.0177(13)$ & $0.2906(45)$ & $2165(2)$\\
\cline{2-5}
&\textbf{Level 4}
& $0.0150(12)$ & $0.2890(45)$ & $3263(3)$\\
\hline
\multicolumn{2}{|c|}{\textbf{MIP}} & $0.0159(1)$ & $0.2833(5)$ & \\
\hline
\end{tabular}
\caption{Logical error rates $p_L$ for different decoders are reported for physical error rates $p=0.05$ and $p=0.15$ for the distance-7 surface code under bit-flip noise. The standard deviation is shown in parentheses.
}
\label{tab:surface}
\end{table}

In Table~\ref{tab:surface}, we summarize the logical error rates and the rank of the moment matrix for the distance-7 surface code.

\section{Honeycomb Color Code}\label{sec:color}

The honeycomb color code is a two-dimensional topological stabilizer code defined on a trivalent, three-colorable lattice, typically the hexagonal (honeycomb) tiling, in which qubits are placed on lattice vertices and each face is assigned one of three colors such that no two adjacent faces share the same color \cite{bombin2006topological, bombin2007exact}. As a CSS code, it associates both an $X$-type and a $Z$-type stabilizer with every face, with each stabilizer acting on all qubits along the boundary of that face. Since the lattice is trivalent and admits a global face-coloring, all $X$-type stabilizers commute with all $Z$-type stabilizers, ensuring a consistent CSS structure and enabling fault-tolerant syndrome extraction.

For each face $f$, the stabilizers are defined as products of Pauli operators around the face boundary,
\[
S_f^{X} = \prod_{v\in\partial f} X_v,
\;
S_f^{Z} = \prod_{v\in\partial f} Z_v,
\]
where $\partial f$ denotes the ordered set of vertices along the boundary of $f$. In the honeycomb tiling, every bulk face has degree six, and thus the corresponding stabilizers have weight~6, while boundary truncations yield stabilizers of smaller weight. The CSS parity-check matrices $H_X$ and $H_Z$ are obtained by assigning one row per face and one column per qubit, with entries indicating stabilizer support, i.e., $(H_X)_{f,v}=1$ if $v\in\partial f$, and likewise for $H_Z$. The three-colorability of the lattice ensures the commutation condition $H_X H_Z^{T} = 0$, satisfying the defining requirement for a CSS code~\cite{bombin2006topological}.

\begin{figure}[htpb]
\centering
\subfigimg[width=0.3\columnwidth]{a}{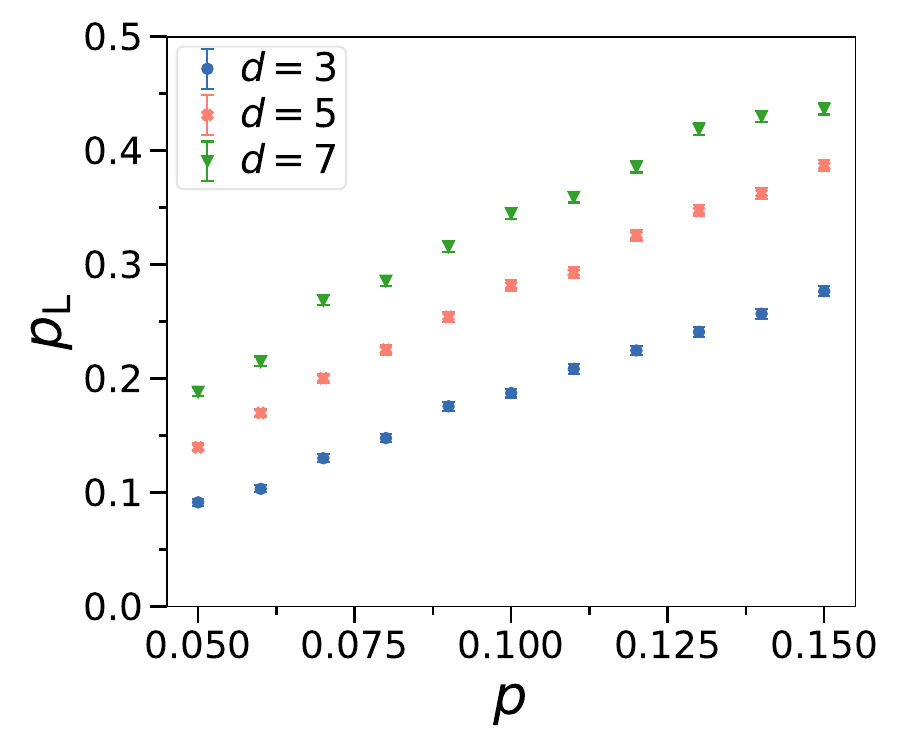}
\subfigimg[width=0.3\columnwidth]{b}{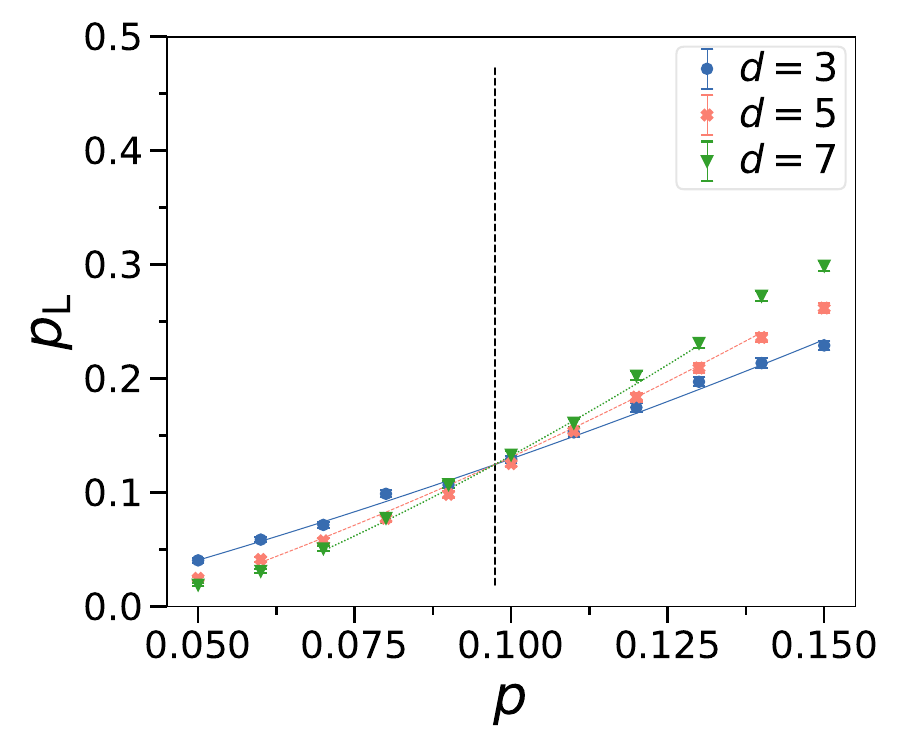}\\
\subfigimg[width=0.3\columnwidth]{c}{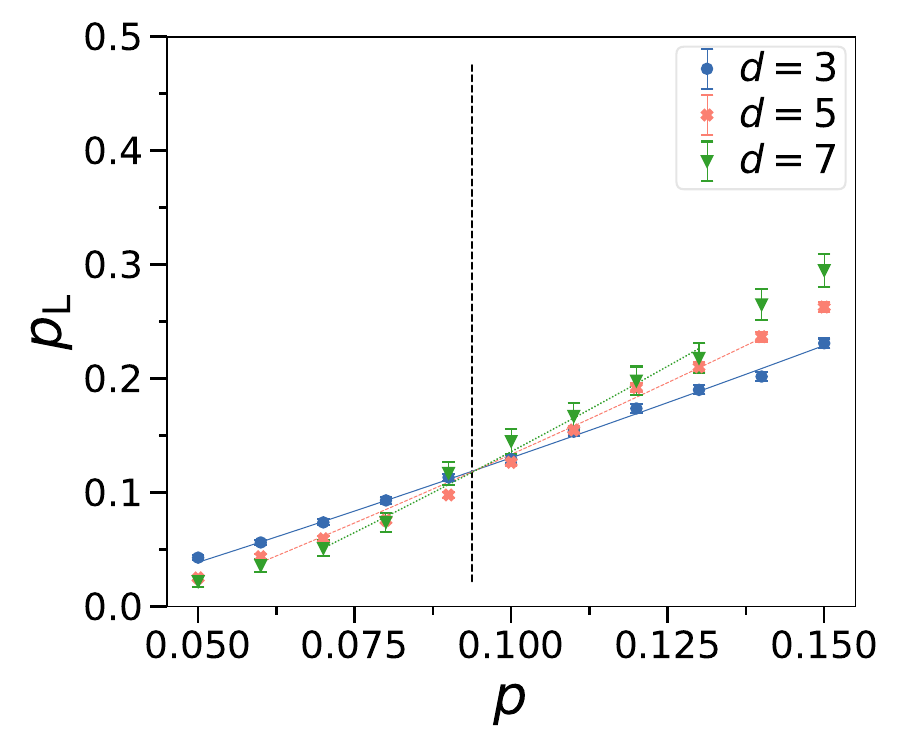}
\subfigimg[width=0.3\columnwidth]{d}{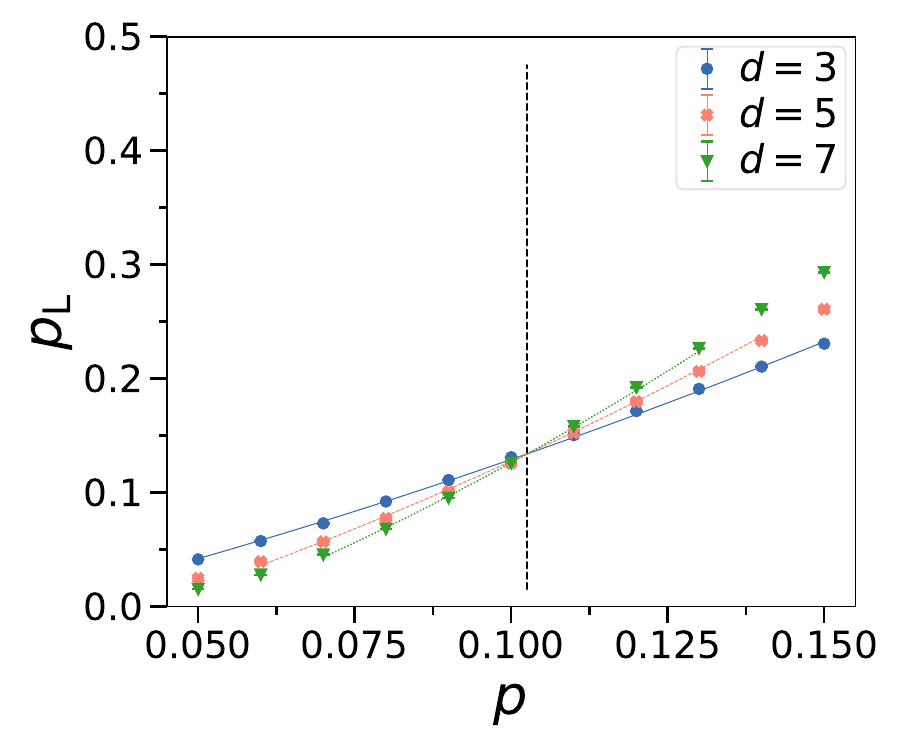}
\caption{Decoding of color code. We show logical error $p_\text{L}$ against physical error $p$ for different code distances $d$. We show SOS hierarchy \idg{a} level 1, \idg{b} level 2, \idg{c} level 3 and \idg{d} MIP decoder. 
Vertical dashed line shows the numerically fit of threshold $p_\text{th}=\{\text{N/A}, 0.095, 0.098, 0.103\}$.
}
\label{fig:colorcode}
\end{figure}

\begin{figure}[htpb]
\centering
\subfigimg[width=0.35\columnwidth]{}{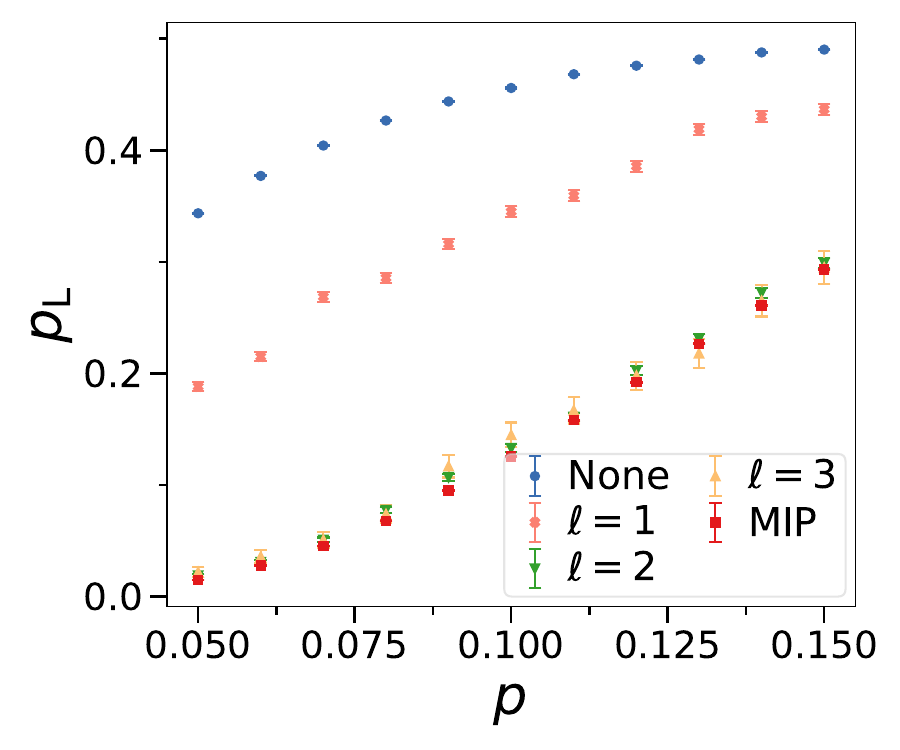}
\caption{Comparison of different SOS levels for decoding of color code. We show logical error $p_\text{L}$ against physical error $p$ for rotated surface code for different types of decoders, where we fix distance $d=7$. We show no decoding at all (blue curve), SOS decoder with level $\ell=1,\dots,3$ (orange, green, yellow, respectively) and the MIP decoder (red).
}
\label{fig:colorcode_comp}
\end{figure}

Fig.~\ref{fig:colorcode} shows the logical error rate $p_\text{L}$ for the color code. 
We also compare the performance for fixed distance $d$ for different decoders in Fig.~\ref{fig:colorcode_comp}. We find that while level $\ell=1$ of our decoder gives only a modest improvement compared to no decoding at all, for $\ell\geq2$ we find performance comparable to the optimal MIP decoder.

\section{Equivalence between classical decoding and MLD}\label{sec:MLD}

In this section, we show that classical decoding and MLD for quantum codes are equivalent. 
To decode a classical code $\codes$, the \emph{maximum a posteriori} (MAP) codeword $x^*$ is the one that was most likely transmitted given that the codeword $y$ was received
\begin{equation*}
    x^* = \arg\max_{x \in \codes} \Pr[x\ \text{transmitted}\ |\ y\ \text{received}].
\end{equation*}

By the Bayes rule and assuming that all of the codewords have equal probability, the MAP codeword is the same codeword $x$ that maximizes the probability $y$ was received, given that $x$ was transmitted
\begin{equation*}
    x^* = \arg\max_{x \in \codes} \Pr[y\ \text{received}\ |\ x\ \text{transmitted}].
\end{equation*}

This is referred to as the \emph{maximum likelihood} (ML) codeword.

Given that we have received the codeword $y$, we define the \emph{log-likelihood ratio} (LLR) $\gamma_i$ of $x_i$ to be
\begin{equation*}
    \gamma_i = \ln \left( \frac{\Pr[y_i | x_i = 0]}{\Pr[y_i | x_i = 1]} \right).
\end{equation*}

The sign of the LLR $\gamma_i$ determines whether transmitted bit $x_i$ is more likely to be $0$ or $1$.
If we consider $\gamma_i$ as the cost of setting $x_i$ to $1$, we can refer to $\sum_i \gamma_i x_i$ as the cost of the codeword $x$.

Theorem $2.1$ of~\cite{feldman2003decoding} states that for any binary-input memoryless channel, the codeword of minimum cost is the ML codeword.

Thus, the classical decoding problem can be formulated as
\begin{equation}
    \label{eqn:MLD_problem}
    x^* = \arg\min \left\{ \sum_{i=1}^n \gamma_i x_i : x \in \field_2^n, Hx = 0 \right\}.
\end{equation}

We now show the formal proof of equivalence between classical decoding and MLD. Suppose the received codeword is $y$ and the corresponding syndrome $s = Hy$.
The error $e$ and the transmitted codeword $x$ are related by $y = x \oplus e$.
This can equivalently be expressed as $e = y \oplus x$.

For every feasible $x$,
\begin{equation*}
    e = y \oplus x \implies He = Hy \oplus Hx = Hy = s.
\end{equation*}
Every feasible $x$ in~\eqref{eqn:MLD_problem} maps to feasible $e$ in~\eqref{eqn:MLE_problem} for syndrome $s$.

Similarly, for every feasible $e$,
\begin{equation*}
    x = y \oplus e \implies Hx = Hy \oplus He = s \oplus s = 0.
\end{equation*}

Thus there is a bijection between $x \in \{ x \in \field_2^n, Hx = 0 \}$ and $e \in \{ e \in \field_2^n, He = s \}$.

Now, consider the objective function of~\eqref{eqn:MLE_problem}
\begin{equation*}
    \sum_i \gamma_i e_i = \sum_i \gamma_i (y_i \oplus x_i).
\end{equation*}
Observe that
\begin{equation*}
    y_i \oplus x_i =
    \begin{cases}
        x_i & y_i = 0 \\
        1-x_i & y_i = 1
    \end{cases}
\end{equation*}
So, we have
\begin{equation*}
    \begin{aligned}
        \sum_i \gamma_i (y_i \oplus x_i) &= \sum_{i: y_i = 1} \gamma_i + \sum_{i: y_i = 0} \gamma_i x_i - \sum_{i: y_i = 1} \gamma_i x_i \\
        &= k + \sum_i \lambda_i x_i
    \end{aligned}
\end{equation*}
where $k = \sum_{i: y_i = 1} \gamma_i$ and $\lambda_i = (-1)^{y_i} \gamma_i$.

Thus, we get
\begin{equation*}
    \arg\min_{e \in \field_2^n : He = s} \sum_i \gamma_i e_i = \arg\min_{x \in \field_2^n : Hx = 0} \sum_i \lambda_i x_i.
\end{equation*}

\section{Lift-and-project hierarchies}\label{sec:lift-and-project}

Lift-and-project hierarchies are systematic methods for tightening LP relaxations of integer programs. 
The idea is to introduce auxiliary variables representing products of the original variables, add constraints in this higher-dimensional ``lifted'' space, and then project back to the original variables. 
These hierarchies provide progressively tighter relaxations that can eventually recover the exact integer solution.

\subsection{Lovasz-Schrijver hierarchy}

The Lovasz-Schrijver~\cite{lovasz1991cones} (LS) hierarchy applies an operator to a convex programming relaxation $\rlxn$ of a binary linear program to produce a tighter relaxation.
There are two versions of this relaxation, where we denote the weaker version by LS and the stronger version by LS+.
In LS we add auxiliary variables and linear inequalities, and denote the projection of this relaxation on the original variables by $N(\rlxn)$.
LS+ further adds SDP constraints, and we denote the projection on the original variables by $N_+(\rlxn)$.

Iteratively applying the operator $N$ (resp. $N_+$) on the basic relaxation $N(\rlxn)$ results in higher levels of the LS (resp. LS+) hierarchy.

For the set $\rlxn \subseteq [0, 1]^n$, we define $cone(\rlxn) \subseteq \mathbb{R}^{n+1}$ as the set of all vectors $(\lambda, \lambda y_1, \cdots, \lambda y_n)$ such that $\lambda \geq 0$ and $(y_1, \cdots, y_n) \in \rlxn$.

We want a solution $(1, y_1, \cdots, y_n)$ to satisfy the conditions $y_i^2 = y_i$.
To achieve this we can introduce $n^2$ new variables $Y_{i,j}$ with the conditions $Y_{i,i} = y_i$ and $Y_{i,j} = y_i \cdot y_j$.
However the latter condition is neither linear nor convex and thus will be approximated by the conditions in the following definition.

\begin{definition}[Protection matrix]
    For a cone $K \subseteq \Rspace^d$, the cone $N(K) \subseteq \Rspace^d$ is defined as follows:
    A vector $y = (y_0, \cdots, y_{d-1}) \in \Rspace^d$ is in $N(K)$ if and only if there is a matrix $Y$ called the protection matrix of $y$ such that
    \begin{enumerate}
        \item $Y$ is symmetric
        \item For each $i \in \{0, \cdots, d-1\}$, $Y_{0, i} = Y_{i, i} = y_i$
        \item Each row $Y_i$ belongs to $K$
        \item Each vector $Y_0 - Y_i$ is an element of $K$
    \end{enumerate}
    Here, $Y_i$ denotes the $i$-th row of $Y$.
    If in addition, $Y$ is positive semidefinite, then $y \in N_+(K)$.
    The $t$-th level of the hierarchy, denoted as $N^t(K)$, is obtained as $N(N^{t-1}(K))$ (respectively, $N_+(N_+^{t-1}(K))$), with the zeroeth level $N^0(K)$ and $N_+^0(K)$ being the same as $K$.
    When $K = cone(\rlxn)$ for $\rlxn \subseteq \Rspace^{d-1}$, the set $N^t(\rlxn)$ is defined as $\{ y \in \Rspace^{d-1}\ :\ (1, y) \in N^t(cone(\rlxn)) \}$.
    The set $N_+^t(\rlxn)$ is defined similarly.
\end{definition}

The level $1$ Lovasz-Schrijver hierarchy for MLD is given by
\begin{equation}
    \begin{aligned}
        \min \quad& \sum_{i=1}^n \gamma_i e_i \\
        \textrm{s.t.} \quad& Y_{0,0} = 1 \\
        & Y_{0,i} = Y_{i,i} = e_i\ i \in \{1, \cdots, n\} \\
        & Y_{i,j} = Y_{j,i}\ \forall\ i,j \\
        & H \cdot Y_i = Y_{0,i} \cdot s\ i \in \{0, \cdots, n\} \\
        & H \cdot (Y_0 - Y_i) = (1 - Y_{0,i}) \cdot s\ \forall\ i \\
        & 0 \leq Y_{i,j} \leq Y_{0,i}\ \forall\ i,j
    \end{aligned}
\end{equation}

\subsection{Sherali-Adams hierarchy}

The Sherali-Adams hierarchy~\cite{sherali1990hierarchy} can be viewed as a strengthening of the LS hierarchy.
In the Lovasz-Schrijver hierarchy, we have the variable $Y_{ij}$ which we wanted to express as $Y_{ij} = y_i \cdot y_j$.
Consider a solution $(1, y_1, \cdots, y_n)$ which is feasible at $LS^{(2)}$.
Then the row $Y_i$ of the protection matrix must also define a feasible solution to the cone version of the relaxation.
The solution $y' = Y_i = (Y_{i0}, Y_{i1}, \cdots, Y_{in})$ must be feasible for $LS^{(1)}$, and so there exists a corresponding protection matrix $Y'$.
Now we would like $Y'_{jk}$ to be a relaxation for $y_i y_j y_k$.
However, the choice of $Y'$ was dependent on the fact that we first chose the row $Y_i$.
If we looked at $Y''$ for the solution $y'' = Y_j = (Y_{j0}, Y_{j1}, \cdots, Y_{jn})$, it need not be true that $Y_{jk}' = Y_{ik}''$.

The Sherali-Adams (SA) hierarchy solves this problem by introducing all the auxiliary variables at once instead of doing this in an inductive manner.
We define a variable $Y_S$ for each $S \in \{1, \cdots, n\}$ with $|S| \leq t+1$.
The idea is that instead of imposing $Y_S = \prod_{i \in S} y_i$, we impose linear conditions imposed by this. 
For each constraint $a^\top y - b \leq 0$ of $\rlxn$, consider sets $S$ and $T$ such that $|S| + |T| \leq t$ and impose $(a^\top y - b) \cdot \prod_{i \in S} y_i \cdot \prod_{j \in T}(1 - y_j) \leq 0$, by requiring that
\begin{equation*}
    \sum_{T' \subseteq T} (-1)^{|T'|} \cdot \left( \sum_{i=1}^n a_i \cdot Y_{S \cup T' \cup \{i\}} - b \cdot Y_{S \cup T'} \right) \leq 0.
\end{equation*}
Each level is a relaxation, since for any $y \in \{0, 1\}^n$ satisfying the initial constraints, $Y_S = \prod_{i \in S} y_i$ defines a valid level-$t$ solution.

The level $\ell$ Sherali-Adams hierarchy for MLD can be formulated as
\begin{equation}
    \begin{aligned}
        \min \quad& \sum_{i=1}^n \gamma_i Y_{\{i\}} \\
        \textrm{s.t.} \quad& Y_{\emptyset} = 1 \\
        & Y_S \geq 0\ \forall\ S\ \text{with}\ |S| \leq t + 1 \\
        & \sum_j H_j \cdot Y_{S \cup \{j\}} = Y_S \cdot s\ \forall\ S\ \text{with}\ |S| \leq t + 1 \\ %
        & 0 \leq Y_{S \cup \{i\}} \leq Y_S\ i \notin S, |S| \leq t
    \end{aligned}
\end{equation}

\section{Comparison of relaxations}\label{sec:comp_rlxn}
\begin{figure}
\centering
\includegraphics{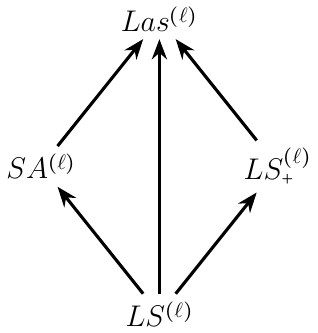}
\caption{Relationship of $\ell$-th level of $SA^{(\ell)}$ (Sheralli-Adams), $\ LS^{(\ell)}$, $\ LS_+^{(\ell)}$ (Lovasz-Schrijver), and $\ Las^{(\ell)}$ (Lasserre) relaxations, where the direction of arrows denotes tighter relaxations~\cite{chlamtac2012convex}. %
}
\label{fig:compare_relaxations}
\end{figure}

Let $SA^{(\ell)}(\rlxn),\ LS^{(\ell)}(\rlxn),\ LS_+^{(\ell)}(\rlxn),\ Las^{(\ell)}(\rlxn)$ denote the feasible sets corresponding respectively to $\ell$ levels of the Sheralli-Adams, LS, LS+ and Lasserre hierarchies, obtained by starting from a basic linear relaxation $\rlxn$ for some binary integer program.
Fig.~\ref{fig:compare_relaxations} compares these relaxations.

These hierarchies compare as follows:
\begin{enumerate}
    \item $LS^{(n)}(\rlxn) = LS_+^{(n)}(\rlxn) = SA^{(n)}(\rlxn) = Las^{(n)}(\rlxn) = \I$, where $\I$ denotes the convex hull of the solutions to the binary integer program over $n$ variables.
    \item For all $\ell \leq n, LS^{(\ell)}(\rlxn) \subseteq LS_+^{(\ell)}(\rlxn) \subseteq Las^{(\ell)}(\rlxn)$, and also $LS^{(\ell)}(\rlxn) \subseteq SA^{(\ell)}(\rlxn) \subseteq Las^{(\ell)}(\rlxn)$.
    Thus, the relaxations provided by the Lasserre hierarchy at each level are the strongest among all the others.
    \item If $\rlxn$ has $n^{O(1)}$ constraints, then $LS^{(\ell)}(\rlxn), LS_+^{(\ell)}(\rlxn), SA^{(\ell)}(\rlxn)$ and $Las^{(\ell)}(\rlxn)$ takes time $n^{O(\ell)}$ to optimize over.
    While it is known that one can optimize efficiently over $LS^{(\ell)}(\rlxn), LS_+^{(\ell)}(\rlxn)$ and $SA^{(\ell)}(\rlxn)$ even if we only assume that $\rlxn$ has a weak separation oracle running in time $n^{O(1)}$, this is not known for $Las^{(\ell)}(\rlxn)$ even with an efficient separation oracle for $\rlxn$.
\end{enumerate}

We refer the interested reader to~\cite{chlamtac2012convex} for further details.

\section{QUBO formulation of MLD}\label{sec:MLE_QUBO}

Since $e_j^2 = e_j, \ j \in \{1, \dots, n\}$, the objective function
\begin{equation}
    \gamma^\top e = \sum_{i=1}^n e_i \gamma_i = \sum_{i=1}^n e_i^2 \gamma_i = e^\top \Gamma e,
\end{equation}
where $\Gamma$ is a $n \times n$ diagonal matrix with entries $\Gamma_{ii} = \gamma_i$ for each value of $i$.

Now consider MLD
\begin{equation}
    \begin{aligned}
        \min \quad& e^\top \Gamma e \\
        \mathrm{s.t} \quad& He = s \\
        \quad& e \in \F_2^n.
    \end{aligned}
\end{equation}

Let us introduce a quadratic penalty term $\xi$ for the constraint $He = s$.
We have
\begin{equation}
    \begin{aligned}
        y &= e^\top \Gamma e + \xi\ (He - s)^\top (He - s) \\
        &= e^\top \Gamma e + \xi\ (e^\top H^\top H e - 2 s^\top H e + s^\top s) \\
        &= e^\top \Gamma e + e^\top (\xi H^\top H) e + e^\top \Delta e + \xi s^\top s \\
        &= e^\top Q e + \xi s^\top s.
    \end{aligned}
\end{equation}
Here $\Delta$ is a $n \times n$ diagonal matrix with entries $\Delta_{ii} = -2 \xi (s^\top H)_i$ for each value of $i$.

Since $s^\top s$ is a constant, we have the following QUBO problem
\begin{equation}
    \begin{aligned}
        \min \quad& e^\top Q e \\
        \mathrm{s.t} \quad& e \in \F_2^n.
    \end{aligned}
\end{equation}

\paragraph*{Arbitrary CSS codes}

Let $e_X \in \F_2^n$ indicate whether or not qubit $j$ experienced bit flip ($X$ or $Y$), and
let $e_Z \in \F_2^n$ indicate whether or not qubit $j$ experienced phase flip ($Z$ or $Y$).

Suppose $H_Z \in \mathbb{F}_2^{r_Z \times n}$ denote the parity check matrix for bit-flip errors, and $H_X \in \mathbb{F}_2^{r_X \times n}$ denote the parity check matrix for phase-flip errors.
Let $s_Z \in \F_2^{r_X}$ and $s_X \in \F_2^{r_Z}$ denote the measured syndromes for the bit-flip errors and phase-flip errors, respectively.

Let
\begin{equation*}
    \begin{aligned}
        \gamma_i^X &\coloneqq \log \left( \frac{1 - p_i^\text{bit}}{p_i^\text{bit}} \right), \\
        \gamma_i^Z &\coloneqq \log \left( \frac{1 - p_i^\text{phase}}{p_i^\text{phase}} \right)
    \end{aligned}
\end{equation*}
for priors $p_i^{\text{bit}} = p_i^X + p_i^Y$, and $p_i^{\text{phase}} = p_i^Z + p_i^Y$.

To decode the bit-flip errors, we solve
\begin{equation*}
    \min \left\{ \sum_i e_{X,i}\ \gamma_i^X\ :\ H_Z e_X = s_Z,\ e_{X, i}^2 - e_{X, i} = 0\ \forall\ i \right\},
\end{equation*}
and to decode the phase-flip errors, we solve
\begin{equation*}
    \min \left\{ \sum_i e_{Z,i}\ \gamma_i^Z\ :\ H_X e_Z = s_X\ e_{Z, i}^2 - e_{Z, i} = 0\ \forall\ i \right\}.
\end{equation*}

By introducing penalty terms $\xi_Z$ and $\xi_X$, we can formulate these problems as
\begin{equation*}
    \min \left\{ e_X^\top Q_X e_X\ :\ e_X \in \F_2^n \right\} \text{ and} \min \left\{ e_Z^\top Q_Z e_Z\ :\ e_Z \in \F_2^n \right\}.
\end{equation*}

Instead of solving these two QUBO problems separately, we can combine these as follows.
Let
\begin{equation*}
    e = 
    \begin{pmatrix}
        e_X \\
        e_Z
    \end{pmatrix}, \quad
    Q = 
    \begin{pmatrix}
        Q_X & 0 \\
        0 &  Q_Z
    \end{pmatrix}.
\end{equation*}

Then we have the following QUBO problem
\begin{equation}\label{eq:CSS_QUBO}
    \min \left\{ e^\top Q e\ :\ e \in \F_2^{2n} \right\}.
\end{equation}

\end{document}